\documentclass[journal,10pt]{IEEEtran}
\usepackage{cite}
\usepackage{amsmath,amssymb,amsfonts}
\usepackage{algorithmic}
\usepackage{graphicx}
\usepackage{multirow}
\usepackage{textcomp}
\usepackage{hyperref}
\usepackage{array}
\usepackage{xcolor}
\usepackage{url}
\usepackage{flushend}
\usepackage{pgfplots}
\usepackage{tabularx}
\usepackage{makecell}

\newcommand{\etal}{\textit{et al. }}

\usepackage[ruled,vlined,linesnumbered]{algorithm2e}
\def\BibTeX{{\rm B\kern-.05em{\sc i\kern-.025em b}\kern-.08em
    T\kern-.1667em\lower.7ex\hbox{E}\kern-.125emX}}
\begin{document}

% \title{}

\author{\IEEEauthorblockN{Shaashwat Agrawal\IEEEauthorrefmark{1},
Sagnik Sarkar\IEEEauthorrefmark{1}, Ons Aouedi\IEEEauthorrefmark{2}, Gokul Yenduri\IEEEauthorrefmark{3}, Kandaraj Piamrat \IEEEauthorrefmark{2}, Sweta Bhattacharya \IEEEauthorrefmark{3}, \\Praveen Kumar Reddy Maddikunta\IEEEauthorrefmark{3}, Thippa Reddy Gadekallu\IEEEauthorrefmark{3}}

\IEEEauthorblockA{
\IEEEauthorrefmark{1}School of Computer Science and Engineering, Vellore Institute of Technology, Vellore, India\\
E-mail:\{shaashwat.agrawal2018,sagnik.sarkar2018\}@vitstudent.ac.in\\
\IEEEauthorrefmark{2} University of Nantes, France \\
E-mail: \{ons.aouedi, kandaraj.piamrat\}@ls2n.fr\\
\IEEEauthorrefmark{3}School of Information Technology, Vellore Institute of Technology, Vellore, India\\
E-mail:gokul.yenduri2020@vitstudent.ac.in,\\ \{sweta.b,praveenkumarreddy,thippareddy.g\}@vit.ac.in\\
}}
		
\title{Federated Learning for Intrusion Detection System: \\Concepts, Challenges and Future Directions}

% \author{Thippa Reddy Gadekallu and Quoc-Viet Pham
 
% \thanks{Thippa Reddy Gadekallu is with School of Information Technology and Engineering, Vellore Institute of Technology, India (e-mail: thippareddy.g@vit.ac.in).}
% \thanks{Quoc-Viet Pham is with Korean Southeast Center for the 4th Industrial Revolution Leader Education, Pusan National University, Busan 46241, Korea (e-mail: vietpq@pusan.ac.kr).}
% }
 
\maketitle

\begin{abstract}
The rapid development of the Internet and smart devices trigger surge in network traffic making its infrastructure more complex and heterogeneous. The predominated usage of mobile phones, wearable devices and autonomous vehicles are examples of distributed networks which generate huge amount of data each and every day. The computational power of these devices have also seen steady progression which has created the need to transmit information, store data locally and drive network computations towards edge devices. Intrusion detection systems play a significant role in ensuring security and privacy of such devices. Machine Learning and Deep Learning with Intrusion Detection Systems have gained great momentum due to their achievement of high classification accuracy. However the privacy and security aspects potentially gets jeopardised due to the need of storing and communicating data to centralized server. On the contrary, federated learning (FL) fits in appropriately as a privacy-preserving decentralized learning technique that does not transfer data but trains models locally and transfers the parameters to the centralized server. The present paper aims to present an extensive and exhaustive review on the use of FL in intrusion detection system. In order to establish the need for FL, various types of IDS, relevant ML approaches and its associated issues are discussed. The paper presents detailed overview of the implementation of FL in various aspects of anomaly detection. The allied challenges of FL implementations are also identified which provides idea on the scope of future direction of research. The paper finally presents the plausible solutions associated with the identified challenges in FL based intrusion detection system implementation acting as a baseline for prospective research.

\end{abstract}


\section{Introduction}

%plan: 
%IDS........

The Internet has greatly expanded in size as well as in complexity over the last decade. At the same time, new types of connected devices have emerged. As a consequence, the network systems are subjected to several security vulnerabilities including intrusions due to the \textcolor{black}{increase in network complexity.} According to~\cite{b5}, intrusions are defined as \textit{"attempts to compromise the confidentiality, integrity, or availability of a computer or network, or to bypass the security mechanisms of a computer or network"}. Therefore, in order to efficiently mitigate different types of attacks, developing an accurate Intrusion Detection System (IDS) is the need of the hour. IDS is an important component for network security~\cite{b2}, which is the first line of defense in the network. %It is a detection system deployed to monitor the network. 
To do so, IDS uses different techniques for detecting anomalies. However, traditional methods such as signature-based detection is becoming less efficient since \textcolor{black}{this technique} uses a signature database to recognize known attacks, and hence new attacks cannot be identified~\cite{b1}.

In recent years, Machine Learning (ML) including shallow models, especially Deep Learning (DL) have advanced considerably and being widely adopted in several domains such as healthcare systems~\cite{b28}~\cite{b32}, computer vision~\cite{b26}, and wireless communication~\cite{b27}. It can provide methods to detect different types of attacks without extensive human-based intervention. For example, IDS using ML techniques learns from traffic (normal/abnormal), implementing the training process of the model. It is implemented on the dataset or the environment to identify pattern relevant to malicious traffic~\cite{b8}. Although these models have been successfully used for IDS, they typically require a central entity to process the data collected from all users in the network. %For example, 
~\cite{b31} found that the prediction accuracy decreases when network scale increases (the more the switches, the less the accuracy), due to high packet loss rate. %Furthermore, there is a wide range of associated constraints which are: (i) the users need to upload their own private data to a central entity for training central model. (ii) IDS requires fast analysis while the centralized processing is time-consuming, and (iii) the IoT devices often collect data of end-users and can lead to exposing their sensitive data.

\begin{figure}[h!]
	\centering
	\includegraphics[width=1.0\linewidth]{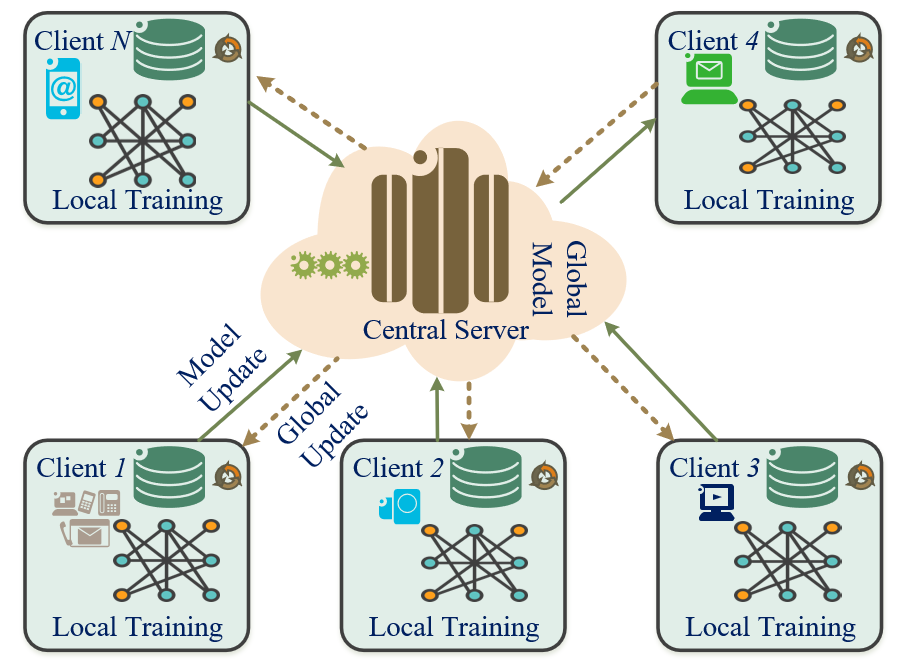}
	\caption{General Federated Learning Architecture.}
	\label{Fig:FL}
\end{figure}

Motivated by the above issues, Google proposed the concept of Federated Learning (FL) for on-device learning and data privacy preservation~\cite{b13,pham2021fusion,mothukuri2021survey}. FL enables the devices to learn in a collaborative way without the need of data sharing with a centralized server. In other words, thanks to its features, ML/DL can be trained across multiple devices and servers with decentralized data over multiple iterations \cite{taheri2020fed,cheng2020federated}. In general, it consists of two major steps: local learning and model transmission which enables attainment of privacy preservation and cost reduction which are expected in  traditional centralized machine learning methods~\cite{b19}. FL is an iterative process where in each round, the whole ML/DL model can be improved. At the beginning of each round, the FL server selects a subset of clients to participate in the learning process and disseminates to them its current global model \cite{mothukuri2021federated}. Upon receiving the global model, each client uses its own data for the local training. Then these clients send back their new model (i.e., the learned parameters) to the FL server for global aggregation (Figure~\ref{Fig:FL}). This process is repeated various rounds until the desired performance is achieved. In summary, a federated learning scenario consists of two main phases: local update and global aggregation. %The local update refers to the clients' model training using their local data. Global aggregation includes collecting the updated model parameters by the server from the clients, aggregating these parameters, and sending back the aggregated parameters to be used in the next iteration. 
This demonstrates the fact that with the help of FL, the clients can benefit from the other clients' data without sending their privacy-sensitive personal data to a central server~\cite{b20}.

In general, according to how the clients' data are distributed, FL can be divided into Horizontal Federated Learning (HFL), Vertical Federated Learning (VFL), and Federated Transfer Learning (TFL). In more details, HFL is an FL approach where datasets on the clients (i.e., devices) have the same features space with different observation. VFL (known as features-based FL) is an FL approach where the data between different domains is used to train the global model. Here, dataset on the clients can have the same observations with different features. In addition to HFL and VFL, there is TFL architecture  proposed in~\cite{b22}, which can be used when the dataset on the devices differ not only in instances but also in features. For further details, we refer the reader to~\cite{b15}.

%IDS with FL.......

The research community has advocated the application of FL for IDS in order to achieve anomaly detection in local devices. Consequently, the introduction of FL for IDS enables DL/ML to provide customized defense mechanisms in several devices. Moreover, it helps to alleviate the computational burden of the central processing server, preserves data privacy, improves the bandwidth utilization and further deals better with a surge of diverse data communications. Multiple number of surveys and tutorials related to FL for computer networks have been published in the recent times~\cite{b23}~\cite{b30}. \textcolor{black}{Although there has been a tremendous growth in the development of FL for IDS, there still exists lack of comprehensive survey that focusing exclusively on the applications of FL for IDS.}
%However, although the development of FL and the importance of the IDS, there is a lack of a comprehensive survey that focuses on the applications of FL for IDS. 

\subsection{Related Work}
\label{RWork}
There exists minuscule number of research publications focused exclusively on providing an overview of the use of FL in IDS. But there does exists quite a number of surveys focused on FL and IDS separately though. Hence in order to fill this gap, the present paper intends to provide a comprehensive survey of the FL technique applied to IDS. It is highly expected that the discussion and exploration could give scientific and technical readers an overall understanding of this domain and foster more subsequent studies on this issue. The earlier efforts are summarized in Table~\ref{Trelated}. In this section, the main contributions of the various works in this field are presented which helps to compare the same with our contributions.

%this paper is the first that investigates the relation between FL and IDSs.
%In the literature, we can find several DL/ML-based intrusion detection surveys. 
One of the recent survey paper ~\cite{b1} presented and compared several deep learning-based IDS works. They proposed a fine-grained taxonomy, which classifies the state-of-the-art of deep learning-based IDS with respect to four different aspects: input strategy, detection strategy, deployment strategy, and performance evaluation strategy.

%In a brief work,
~\cite{b3} reviewed only seven deep learning-based solutions for IDS. \cite{b4} presented a comprehensive discussion of using Random Forest models for developing an efficient IDS. In addition,~\cite{b6} proposed a survey on SDN based network intrusion detection system using ML/DL approaches. \cite{b7} presented an overview of intrusion classification algorithms where various ensemble and hybrid techniques were examined, considering both homogeneous and heterogeneous types of ensemble methods. \cite{b8} proposed a survey of shallow and deep networks intrusion detection systems and explained the role of feature selection in the classification and training phase of Machine Learning for IDS.

%.................

%This survey extends the previous surveys by focusing on the utilization of Federated Learning in IDSs.
\begin{table*}[ht]
\begin{center}
\caption{Summary of existing surveys, related to ML/DL learning, FL, and IDSs. The symbol (\checkmark) indicate a publication is in the scope of a domain, \textcolor{black}{x} marks papers that do not cover that area.}
\label{Related}
\begin{tabular}{|c|p{10cm}|c|c|c|c|c|}
\hline
\multirow{2}{*}{\textbf{Ref}} & 
\multirow{2}{*}{\textbf{Brief summary}} & 
\multicolumn{4}{c|}{\textbf{Scope}}\\
\cline{3-6}
 && Shallow ML methods & DL & FL & IDS\\
 \hline
 \cite{b16} & A comprehensive deep learning survey. & \textcolor{black}{x} & \checkmark & \textcolor{black}{x} & \textcolor{black}{x} \\
 \hline
 \cite{b18} & A recent survey on deep learning. & \textcolor{black}{x} & \checkmark & \textcolor{black}{x} & \textcolor{black}{x} \\
 \hline
 \cite{b17} & A tutorial on deep learning & \textcolor{black}{x} & \checkmark & \textcolor{black}{x} & \textcolor{black}{x} \\
 \hline
\cite{b1} &  A survey on the proposed deep learning-based IDSs. & \textcolor{black}{x} & \checkmark & \textcolor{black}{x} & \checkmark\\
\hline
\cite{b3} & An overview of anomaly detection methodologies have been introduced with the topics of data reduction, dimensionality reduction, classification, as well as a group of deep learning techniques. & \checkmark & \checkmark & \textcolor{black}{x} & \checkmark \\
 \hline
\cite{b4} & A comprehensive survey on IDSs using Random Forest as a features selection method and as a classifier. &  \checkmark & \textcolor{black}{x} & \textcolor{black}{x} & \checkmark \\
\hline
\cite{b6} & A brief survey on SDN based network intrusion detection system using machine and deep learning approaches. & \checkmark & \checkmark & \textcolor{black}{x} & \checkmark \\
\hline
\cite{b9} & An overview of various deep learning approaches. & \textcolor{black}{x} & \checkmark & \textcolor{black}{x} & \textcolor{black}{x} \\
\hline
\cite{b7} & An overview of intrusion classification algorithms, based on ensemble and hybrid classifiers. & \checkmark & \textcolor{black}{x} & \textcolor{black}{x} & \checkmark\\
\hline
\cite{b15} & An introduction to the basic concept, architecture, and techniques of federated learning, and discussion of its potential in various applications. &  \textcolor{black}{x} & \textcolor{black}{x} & \checkmark & \textcolor{black}{x} \\
\hline 
\cite{b10} & A survey on deep learning and its applications. & \textcolor{black}{x} & \checkmark & \textcolor{black}{x} & \textcolor{black}{x} \\
\hline
\cite{b8} & A survey of shallow and deep networks for intrusion detection systems. & \checkmark & \checkmark & \textcolor{black}{x} & \checkmark\\
\hline
\cite{b11} &  A comprehensive study of Federated Learning (FL) with an emphasis on enabling software and hardware platforms, protocols, real-life applications and use-cases. & \textcolor{black}{x} & \textcolor{black}{x} & \checkmark & \textcolor{black}{x}\\
\hline
\cite{b12} &  An overview on integrating FL with Industrial Internet of Things in terms of privacy, resource and data management. & \textcolor{black}{x} & \textcolor{black}{x} & \checkmark & \textcolor{black}{x} \\
\hline
\cite{b14} & An introduction to FL along with general applications and challenges. & \textcolor{black}{x} & \textcolor{black}{x} & \checkmark & \textcolor{black}{x} \\
\hline
\cite{b19} & An introduction to FL and its related technologies & \textcolor{black}{x} & \textcolor{black}{x} & \checkmark & \textcolor{black}{x} \\
\hline
\cite{b21} & A comprehensive study concerning FL's security and privacy aspects. & \textcolor{black}{x} & \textcolor{black}{x} & \checkmark & \textcolor{black}{x} \\
\hline
\cite{b23} & A survey on FL in mobile edge networks. & \textcolor{black}{x} & \textcolor{black}{x} & \checkmark & \textcolor{black}{x} \\
\hline
\cite{b24} & A survey on the FL challenges and the main approaches that employed FL to solve networking domain issues.  & \textcolor{black}{x} & \textcolor{black}{x} & \checkmark & \textcolor{black}{x} \\
\hline
\textbf{Our work} & \textbf{A comprehensive survey of federated learning for intrusion detection systems} & \checkmark & \checkmark & \checkmark & \checkmark \\
\hline
\end{tabular} 
\label{Trelated}
\end{center}
\end{table*}

\subsection{Contributions}

This work aims to provide a survey on FL for IDS in terms of achievements and challenges, which makes it stand out in comparison to the previous surveys. In brief, the contributions of this work can be summarized as follows:

\begin{itemize}
    \item It discusses the role of FL in intrusion detection.
    \item It describes how FL can help centralized ML/DL approaches to provide an efficient intrusion detection solution.
    \item It reviews the approaches and technologies that applied ML/DL/FL for intrusion detection.
    \item It highlights open research challenges and outlines possible future research directions in order to help the research community for designing efficient future solutions.
    %\item .......
\end{itemize}

\subsection{Paper Organization}

\begin{figure*}[ht]
\centering
\includegraphics[width=12cm, height=6cm]{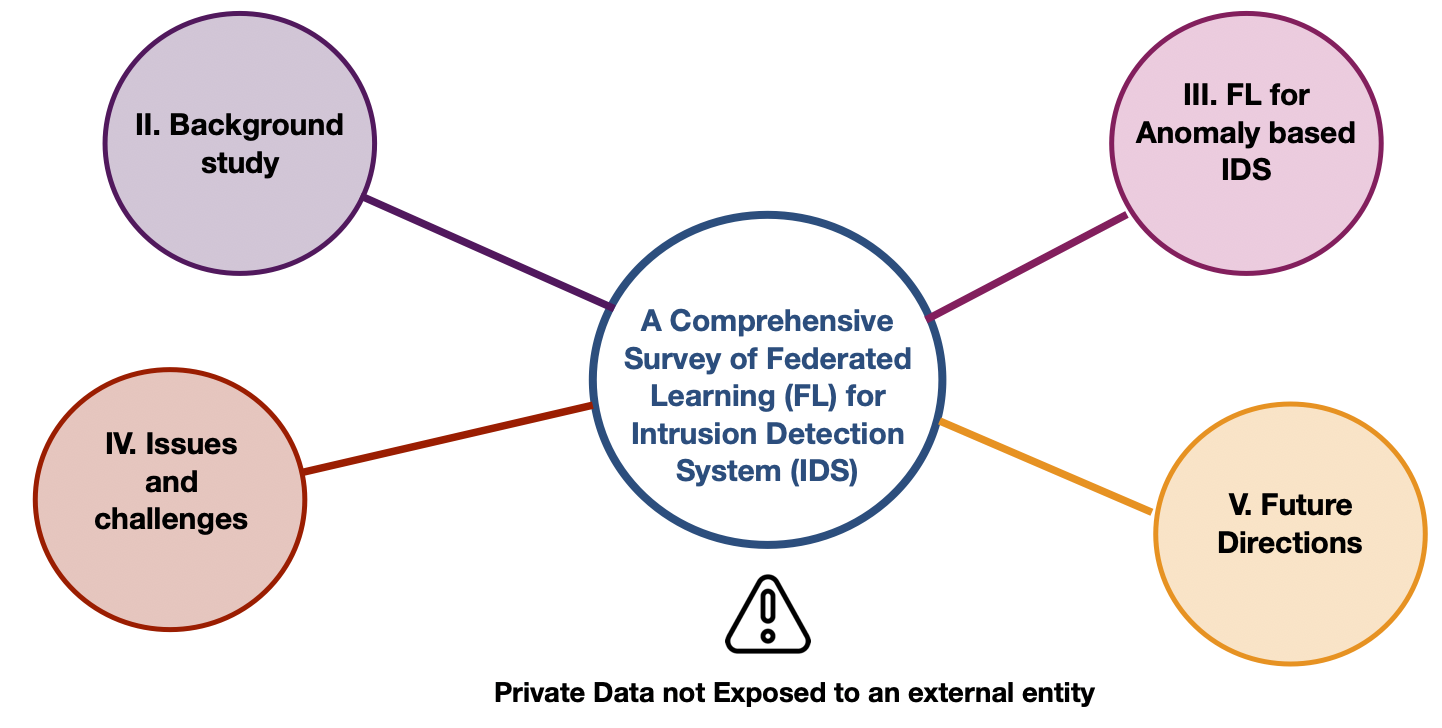}
\vspace{-0.2cm}
\caption{Conceptual map of the paper}
\label{SMap}
\end{figure*}

As illustrated in Figure~\ref{SMap}, the rest of this paper is organized as follows. In section~\ref{back} we give a summary of IDS, as well as the role of machine learning and artificial intelligence (ML/AI) in anomaly intrusion detection. The main purpose of section~\ref{rev} is to provide a comprehensive survey of FL applications with anomaly-based Intrusion Detection Systems. This survey also highlights open challenges and issues of using FL for IDSs as presented in section~\ref{chall}. Next, section ~\ref{FD} discusses future research directions for FL-based intrusion detection. Finally, section~\ref{conc} outlines the conclusions of this paper. For the purpose of better comprehension, the definitions of the abbreviations used in the paper are summarized in Table \ref{Abrev}.

\begin{table}[!htbp]
\centering
\caption{List of abbreviations}
\begin{tabular}{|p{1.5cm}|p{6cm}|}
  \hline
  \textbf{Acronym} &\textbf{Definition}\\
  \hline
  ML & Machine Learning\\
  \hline
  DL & Deep Learning\\
  \hline
  AI & Artificial Intelligence\\
 \hline
  FL & Federated Learning\\
 \hline
  DNN & Deep Neural Network\\
 \hline
 CNN & Convolution Neural Networks\\
 \hline
 LSTM & Long Short Term Memory\\
 \hline
 ADS & Anomaly Detection Systems\\
 \hline
 IDS & Intrusion Detection Systems\\
 \hline
 IIoT & Industrial Internet of Things\\
 \hline
 NIDS & Network Intrusion Detection Systems\\
 \hline
 HIDS & Host Intrusion Detection Systems\\
 \hline
 DAD & Deep Anomaly Detection\\
 \hline
 GAN & Generative Adversarial Network\\
 \hline
 MLP & Multi-Layer Perceptron\\
 \hline
 DDoS & Distributed denial of service\\
 \hline
 MITM & Man in the Middle\\
 \hline
 VPN & Virtual Private Network\\
 \hline
 WSN & Wireless Sensor Network\\
 \hline
 WCN & Wireless Communication Network\\
 \hline
 BNN & Binarized Neural Networks\\
 \hline
 AFL & Asynchronous Federated Learning\\
 \hline
 STIN & Satellite-Terrestrial Integrated Networks\\
 \hline
 ENN & Encrypted Neural Network\\
 \hline
 DRL & Deep Reinforcement Learning\\
 \hline
 DDoS & Distributed denial of service\\
 \hline
 QoS & Quality of Service\\
 \hline
\end{tabular}
\label{Abrev}
\end{table}

\section{Background Study}
\label{back}
Ensuring reliable and trusted information exchange between different entities requires a high degree of protection. IDS assures resilient protection over distinct anomalies. It investigates numerous types of attacks, identifies malicious patterns, and supports administrators in tuning, organizing, and implementing effective controls \cite{mohammadi2021comprehensive,wu2020network}. This section gives a summary of IDS, as well as the role of machine learning and artificial intelligence (ML/AI) in anomaly detection. It also discusses the motivation for the use of federated learning for anomaly-based IDS.

\subsection{Intrusion Detection Systems}
An intrusion detection system (IDS) is a hardware or software system that tracks a network for unauthorised behaviour or policy breaches \cite{bhattacharya2020novel, alazab2016spam}. The IDS logs the details related to the intrusion, initiates warnings and takes necessary mitigation or corrective measures in case an intrusion is detected.
\subsection{Types of IDS}
Host Intrusion Detection System (HIDS) and Network Intrusion Detection System (NIDS) are two types of IDS \cite{9108270}. HIDS is in-charge of monitoring a system internally, including connections to log files, user actions, and so on. NIDS, on the other hand, examines both incoming and outgoing traffic between network nodes.
\subsubsection{Host Intrusion Detection System (HIDS)}
HIDS maintains a record of system events to indicate occurrence of unusual activities and stays updated with device artifacts, operations, memory regions \cite{lee2021towards}. \textcolor{black}{HIDS faces issues pertaining to tampering attacks and the use of anti-tampering mechanism makes its adaption reliable}. A host-based IDS is not an ideal solution on its own. It suffers from severe disadvantages such as high consumption of resources that harms host's performance. Such attacks that may not be exposed unless they penetrate the host.
\subsubsection{Network Intrusion Detection System (NIDS)}
NIDS are deployed at a predetermined location throughout the network to scrutinize traffic from all connected networks \cite{wu2020network}. It interprets all the traffic that passes through the sub-net and based on the comparison with anomalies library, an intrusion is identified.
\textcolor{black}{During the process of identifying anomalies, NIDS remains undetected while monitoring the traffic }. There exists issues in scrutinizing every packet during heavy traffic failing to detect an attack at that time. The problem with the identification of virtual private networks (VPN) and understanding encrypted data add to the list of challenges.
\subsection{Types of Intrusion Detection methods}
There are primarily two approaches for IDS, signature based detection and anomaly based detection \cite{KILINCER2021107840}.
\subsubsection{Signature-based intrusion detection}
Signature-based intrusion detection employs pattern recognition methods to detect attacks, and they are often known as Knowledge-based Detection approaches because they keep track of patterns from past attacks. Based on the information pattern or signature, matching strategies are applied to compare known patterns from the new data. A warning signal would be activated if a pattern looks similar to a previously existing pattern. The ultimate aim is to use a library of known signatures from past attacks to identify the intrusion. Signature-based intrusion detection faces a major challenge; if an intruder uses a new attack that is not already existing in the library, the method will fail to identify the attack. This type of attack is also known as a zero-day attack.
\subsubsection{Anomaly-based intrusion detection }
Anomaly-based intrusion detection overcomes the challenges elevated by signature-based intrusion detection. The detection of intrusion is based on the behavior of the user. A normal model of the system's behavior is generated using statistical-based and other approaches. \textcolor{black}{ The difference between actual and predicted behavior is considered an anomaly}. However, anomaly-based intrusion detection has various limitations.  As an example, it cannot recognize encrypted packets and thus leaves an opportunity for attack. Moreover, the creation of a normal model for enormous dynamic data is extremely challenging which leads to false alarms.

\subsection{Recent trends in IDS based anomaly detection}

As a part of this section, \textcolor{black}some of the achievements of ML/DL in solving recent real-world intrusion detection in various applications have been discussed. 

To safeguard the train–ground communication systems, the authors of the study in \cite{9359546} used ML algorithms, such as random forest and gradient boosted decision tree algorithms, in the primary layer. They used state observer in the second layer to detect abnormalities in the physical state of train operation. A full intrusion detection result is achieved when both layers are combined. This approach uses only two ML techniques, and there is a requirement for more contemporary methods to be applied to enhance the effectiveness of intrusion detection. \textcolor{black}{The authors in \cite{9144275} proposed a two-stage approach to protect Advanced Metering Infrastructure (AMI) from cyber attacks. Support Vector Machines (SVM) is used to identify suspicious behaviors within the meter . The Temporal Failure Propagation Graph (TFPG) is used to generate attack routes that identify possible cyber-attacks, where the proposed pattern recognition algorithm is used to determine the similarity between a newly identified abnormal event and pre-defined cyber attacks}. However, it was designed for a precise type of data, and adaptation of the proposed technique to all types of networks is not feasible. 

Gated Recurrent Unit (GRU)-based Recurrent Autoencoder (RAE) is used by the authors \cite{9211565} to propose a framework named INDRA to identify anomalies in controller area network (CAN) bus-based automotive embedded systems under a variety of different attack scenarios. \textcolor{black}{The adaption of the proposed IS metric for measuring the deviation of the prediction signal from the actual input still remains a concern.} Smart Grids provide reliable energy to homes and industries with the help of SCADA (Supervisory Control and Data Acquisition) systems. The SCADA systems are vulnerable to cyber-attacks because they are interconnected and can be accessed remotely. To solve this issue, the authors in  \cite{9236652} used the gradient boosting method for feature selection along with decision-tree-based algorithms to classify the various attacks and normal events. There is a requirement for more contemporary methods to be applied to enhance the effectiveness of the framework. 

Several deep learning techniques were adopted by Tian et al.\cite{8821336} for proposing web attack detection systems. This system is built to safeguard multiple web applications, located across multiple sites in a distributed environment of Edge of Things (EoT).
Although such system is capable of working in real-time, it still needs to adapt better deep learning algorithms for enhanced performance. Farivar et al.\cite{8917652} used intelligent variable structure control to develop a new technique for estimating and compensating attacks initiated in the forward link of nonlinear Cyber-Physical Systems (CPS). A mixture of nonlinear control and artificial neural networks is used in the proposed technique as it is limited to \textcolor{black}{CPS.} and Industrial IoT only.

Considering the review of the recent applications, it is understood that the techniques used have significant contributions, potential and merits undoubtedly.  But the adaptability of these techniques in case of other similar applications remains a concern.

\subsection{Motivation to adapt FL}

Even though ML and DL are making significant contributions to real-world problem solving, \textcolor{black}{they have a wide range of constraints which are:}

\begin{itemize}
    \item \textcolor{black}{The users need to upload their own private data to a central entity for training the central model.}
    \item \textcolor{black}{The performance decreases while network scale increases, and introduces a single point of failure that could compromise the integrity and Quality of Services (QoS).}
    \item \textcolor{black}{IDS requires fast analysis while the centralized processing is time-consuming.}
    \item \textcolor{black}{The IoT devices often collect data of end-users and can lead to exposing their sensitive data.}
    \item \textcolor{black}{Collecting the data in the 5G/6G network is a burdening and costly task since the network consists of highly diversified data types (e.g. text, audio, video, AR/VR).}
\end{itemize}

\textcolor{black}{It can thus be stated that FL as a technology has the potential to solve various issues. It provides data privacy and ubiquitous intelligence, seamless, flexible, scalable, high-performance communication, ultra-low latency and high-energy efficient networking to support highly dynamic time-critical applications. It is due to the existence of these features that FL provides new opportunity for ML/DL algorithms to be applied more efficiently in IDS. }

%, the benefits of FL compelled us to adapt it for anomaly detection.  The following are some of these motivations. FL enables local model training at edge devices and transferring local models to a centralized server rather than transferring sensitive data, in order to create a global model. It aids in the detection of heterogeneous anomalies, provides privacy of decentralized learning methods and data, enables security of low-power IoT devices, and the reduction of processing delays across various communication networks.

\section{FL for Anomaly based IDS}
\label{rev}

In this section, the state-of-the-art review on FL for the use of anomaly-based IDS is exhibited. The practicality of FL architecture for the implementation of IDS is explored through key terms such as privacy, communication, data storage, and efficiency. A summary of the same is provided in table \ref{tab:review}.

\subsection{FL architecture for deployment of IDS}

% Network, Hub Based IDS
% Centralised, Distributed, Peer-to-Peer IDS

\begin{figure*}[h!]
	\centering
	\includegraphics[width=1.0\linewidth]{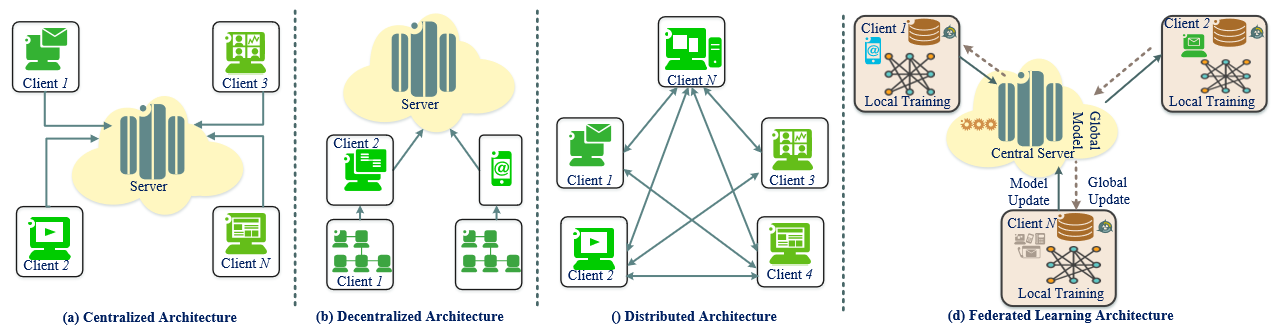}
\caption{Various Architectures for Deployment of IDS.}
\label{fig:arch}
\end{figure*}

The efficiency of any IDS is derived from the precise selection of the deployment architecture and the type of IDS used. There are three major types of IDS deployment architectures that are implemented based on the requirements of the system: Centralized architecture, Distributed architecture, and Decentralized architecture. Centralized deployment of IDS is usually done for small networks with low scalability. This is because this architecture demands the transmission of data to a central server \textcolor{black}{in order to train the ML model for IDS}. Irrespective of the bandwidth of the network, upon increasing the number of devices, the server tends to become a bottleneck for the whole network. Decentralized and Distributed architectures deploy multiple IDS for active detection of attacks. They encompass inter-agent communication and hierarchical decision-making for better intrusion detection but are limited due to the localization of data. Work done in \cite{ferdowsi2019generative} overcomes data localization by implementing distributed Generative Adversarial Networks (GANs) that can monitor intrusions in their systems as well as neighboring systems. This literature is able to improve the distributed working of an IDS but the process may introduce latency in the network. The introduction of federated architecture can overcome the majority of the shortcomings of current IDS architectures. Different IDS deployment types are shown in Fig \ref{fig:arch}.

FL provides a server-client architecture that involves computation at the server end (i.e., model aggregation) as well as the client end (i.e., model training). This style of work division prevents the server from becoming the bottleneck while taking advantage of edge computation. Rey \etal  \cite{rey2021federated} proposed FL frameworks for both Multi-Layer Perceptron (MLP) and autoencoder models in IDS. They conducted experiments using centralized, distributed and FL architectures followed by statistical comparison between them. Their output shows the superiority of FL, which secures higher data utilization. Work done by Rahman \etal \cite{rahman2020internet} evaluates the efficiency of federated architecture in IDS through NSL-KDD dataset \cite{tavallaee2009detailed}. The evaluation is done considering various real-life scenarios and intrusion attacks. A final comparison is made between FL, centralized and self-learning approaches using the obtained experimental results. FL outperformed the other approaches in almost all training rounds. Apart from efficient work division and data utilization, federated architectures are also highly scalable. The larger the federated network, the greater will be its intrusion detection capability.

Depending on the application and user demands, IDS can be deployed in large or small networks (Network Intrusion Detection Systems), on individual host systems (Host Intrusion Detection Systems), or even based on specific protocols (Protocol-based Intrusion Detection Systems). The latency caused by the implementation of a single IDS on huge networks can affect packet transmission rates drastically. A single Network Intrusion Detection System (NIDS) may not be able to handle intrusion attacks through edge devices whereas a Host Intrusion Detection System (HIDS) may not be able to protect the host from indirect intrusions through the network. \cite{8284655} reviews the various types of intrusion detection systems (i.e., NIDS, HIDS, etc.), their data sets, and various limitations. To overcome these limitations and avoid the increase of network latency, IDSs may employ a federated form of deployment. Federated intrusion detection systems are assisted by the size of the network and tend to maximize work division and throughput.

\subsection{FL for Heterogeneous Anomaly Detection}

% Multi-task Learning
% Learning from Different types of traffic 
% Learning from different user's data

In realistic scenarios, anomaly-based IDS tend to produce high amount of false alarms caused by insufficient and unlabelled data, inefficient algorithms, etc. Intrusion attempts by attackers can be modified to avoid IDS by traffic masking, network steganography \cite{9133574}, and other IDS bypassing methodologies. It is pertinent to have sufficient resources for the practical deployment of IDS with low false alarms and high precision. A global dataset with constrained heterogeneity and limited size cannot be used as intrusion attempts are getting trickier and harder to detect. Although distributed IDS train on richer data, they are limited by data size. Hence, FL can be used to solve the problem of insufficient data. It allows access to data available at all client devices which at any instance may suffice the expectations of real-time protection for that network (i.e., larger network implies better protection through the use of bulkier data). Li \etal \cite{li2020enhancing} propose a disagreement based semi-supervised learning for collaborative IDS. The semi-supervised learning algorithm uses three trained classifiers to label the unlabelled data based on majority voting. Each time data is labeled, the models are updated as well. This algorithm utilizes FL for tackling data size limitations and disagreement-based semi-supervised learning for handling unlabelled data. It can provide better results than supervised ML algorithms in terms of prediction accuracy as well as a lesser number of false alarms. 

\begin{figure*}[h!]
	\centering
	\includegraphics[width=1.0\linewidth]{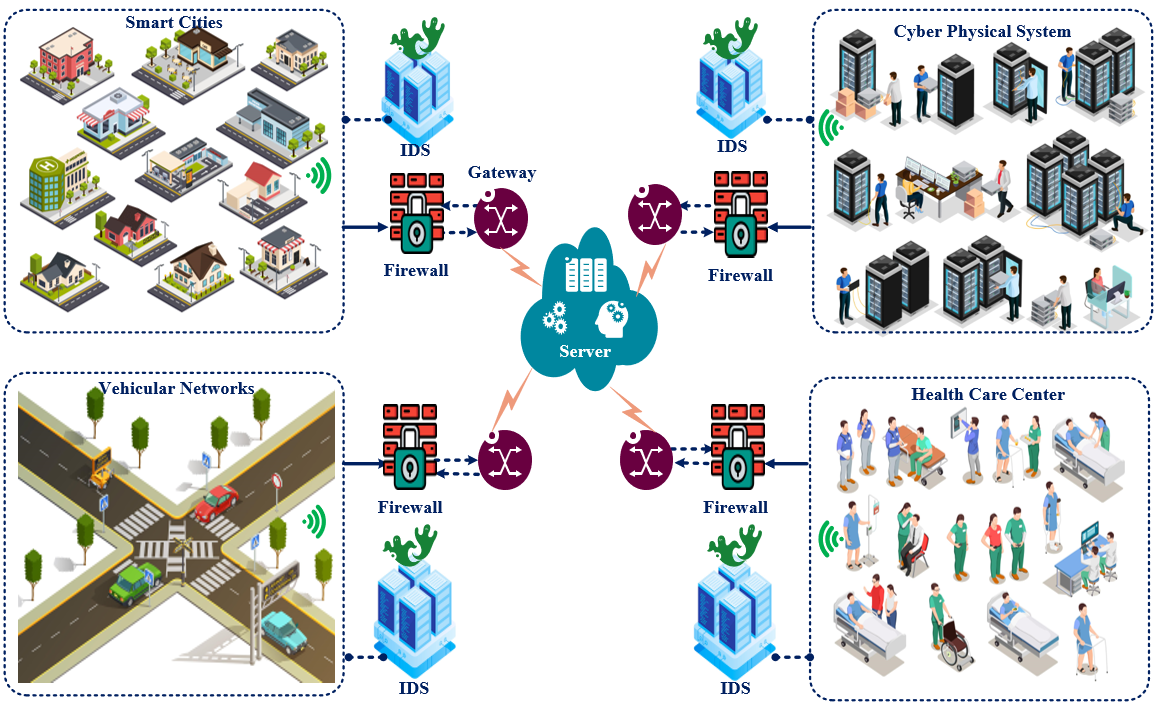}
\caption{FL for heterogeneous Anomaly detection systems based on different environments and applications}
\label{fig:diff}
\end{figure*}

In case an IDS is to be deployed in a multinational company to protect its assets against intrusions, it would be difficult or practically impossible to manage the protection of each node of its network. In such huge networks, each team and the department have different data, communication, and bandwidth requirements resulting in vast dissimilarity in traffic flow properties. Even inside a department, each employee consume different amounts of data. To me more specifice, on certain days it would consume more and and lesser on the others. Fig \ref{fig:diff} depicts the heterogeneity of traffic in IDS deployments as well various applications of FL for IDS. This divergence in data is tedious to be accumulated in a single centralized model. The heterogeneity of data could cause the anomaly detection model to even fail. To handle this vast heterogeneous data, the complexity of federated architecture could be utilized. In \cite{briggs2020federated}, a detailed experimental study is conducted regarding the usefulness and implications of using this data in FL. It conducts experiments on state-of-the-art FL algorithms over heterogeneous and non-IID (Independent and Identically Distributed) data. Although FL shows superiority over centralized architectures, it is still challenged by the non-IID nature of data. 

Another optimization approach in contrast to traditional DL training architectures is provided by \cite{zhao2019multi}. It takes advantage of the heterogeneity of real-time intrusion data to implement the privacy-preserving federated architecture. More importantly, using the same data, a multi-task deep neural network in federated learning (MT-DNN-FL) can perform multiple anomaly detection tasks (i.e., VPN traffic recognition, traffic classification) using a single model. This not only conserves communication energy and training cost but also outperforms several single-task learning methods. 

%--> [ons] I need to do it tonight.
\subsection{FL for DDoS attack Detection}

Distributed Denial of Service (DDoS) attacks acts as one of the major threats to the Internet. The purpose of DDoS is to overload and exhaust the target (i.e., the victims) system by using different computer systems to launch a coordinated DoS attack \cite{criscuolo2000distributed}. Consequently, DDoS attack detection exists as an important task for guaranteeing the QoS of the end-users. In this context, FL has started to attract the attention of researchers for DDoS attack. However, the use of FL for IDS is still at a nascent stage.

For example, \cite{li2020fleam} proposed FL empowered mitigation architecture (FLEAM) in the Industrial Internet of Things (IIoT). FLEAM uses FL, fog, and cloud in handling large volume of DDoS attacks. Specifically, each fog node downloads a training model locally, retrains it and uploads its model parameters to the cloud. The results demonstrates that the mitigation delay is about 72\% lower, and the accuracy is 47\% greater on average than classic solutions.

\subsection{FL for privacy preserving data handling}

The core principle of an IDS is to uphold the confidentiality, integrity, and availability of the information against outside intrusions \cite{rehman2021canintelliids,srivastava2021ensemble}.  Training an anomaly detection statistical model requires huge amounts of data and collecting that data may cause a direct breach in privacy. In traditional Deep Anomaly Detection (DAD) Systems, two important limitations against preserving privacy are identified: presence of large amounts of unprotected data in a central location (i.e., server) and transmission or collection of said data through vulnerable channels. 

\subsubsection{Data at Rest}

% Collaborative Learning: central server and FL
Information theft is one of the leading causes of cyber attacks and intrusion attacks in any system or network. Irrespective of the level of security measures, no system is completely safe, and gathering data in a central location for the training of DAD systems can easily attract intrusion attempts. Data stealing attacks become easier when all the data is located in a single address. Common cloud intrusion attacks like wrapping, flooding, browser malware injection, and honey potting have been identified in \cite{devi2020appraisal}. The work analyzes such attacks in detail and describes how to deal with them. Vishal \etal \cite{vishal2018soaice} simulated such attacks in a cloud-based environment to exploit its security using DDoS attack, jamming attack, etc. To avoid vulnerability exploitation due to these attacks, a FL approach is to be utilized for DAD systems where the data is distributed. 

The data in a federated approach is only available in the client edge devices. Unlike global datasets, federated data cannot be traced or pinpointed, its location and availability are altered dynamically due to the continuous generation of new data in an edge device. Training DAD systems in such heterogeneous environments drastically improve the confidentiality of the data as well as the accuracy of intrusion detection systems. \cite{wang2021towards} proposes a novel Federated Anomaly Detection Systems (ADS) using deep reinforcement learning that utilizes local client data effectively to train IDS. It introduces a privacy leakage degree based on leakage of sensitive and non-sensitive information to identify abnormal users and isolate them. This methodology not only prevents privacy leakage but also attains high anomaly detection accuracy in a hierarchical federated setup. Another major advantage of using federated distribution instead of centralized architecture is that incidental intrusions in one or even several client systems does not affect the system as a whole. The corruption of one or several clients could have an impact on the global model but in realistic scenarios, consisting of a minimum of 10000 clients, the influence is negligible. Similar to \cite{wang2021towards}, \cite{chen2020intrusion} introduces a new algorithm FL-based Attention Gated Recurrent Unit (FedAGRU) for federated training of IDS that includes an attention mechanism for efficient selection of important client updates. FedAGRU was able to get relatively better accuracy and robustness to unimportant and abnormal clients.  

\subsubsection{Data in Motion}

Intrusion detection information usually includes protocol type, service type, source IP address, destination IP address, etc. Transferring these data through open communication channels or even vulnerable routes can be fatal to the users (i.e, source) as well as the server (i.e., destination). Packet transmissions can easily be monitored and altered using Man in the Middle (MITM) attacks. Unlike stationary data, data in transit is not protected by the device's security measures. Thus transferring IDS training data from client devices to a central location is not feasible. Various encryption standards \cite{boulemtafes2020review} as well as compressed learning \cite{zisselman2018compressed} methodologies can be used to avoid outsiders from accessing the information even though the covertness of the documents have been breached. Although these algorithms have been proven useful, they increase the communication and computation overhead. An FL approach, on the other end, trains DL models in local client devices and communicates the same to the central server. A DL model does not represent the data it trains on and hence is hard to attack. As compared to encryption/compression and huge amounts of data communication, FL proves to be very efficient. \cite{cetin2019federated} evaluates such a federated approach for IDS to overcome privacy concerns for sharing sensitive data. It uses wireless communication in a simulated environment to detect intrusions without breaching privacy. Experiments on the Aegean Wi-Fi Intrusion Dataset (AWID) are conducted in this work with a good output accuracy of about 98\% is attained for most classes in an FL setup. 

Although difficult, a model may be able to leak some information about the source data it was trained on using model reverse engineering processes \cite{gongye2020reverse} or back-door attacks on DNNs \cite{li2020invisible}. To prevent these, further research on preventing such reverse engineering attacks \cite{8839491} and back-door attacks \cite{8835365} have been done on DL architectures. Apart from the above developments, work on FL architectures including encryption algorithms, secure communication protocols, and other federated learning-based strategies have also been done to improve the security of model transmissions. Li \etal \cite{li2020deepfed} proposed a novel algorithm DeepFed that uses a federated learning approach using Convolutional Neural Networks (CNNs)  and Gated Recurrent Unit (GRU) deep learning architectures for IDS in industrial cyber-physical systems. To prevent aforementioned attacks on deep learning models, DeepFed also uses a  Paillier cryptosystem-based secure communication protocol that can protect the privacy of model parameters. Using the cryptosystem, a pair of public and private keys are generated. The public key encrypts the local model weights and parameters using the public key after training. The encrypted model parameters are aggregated and sent back to the clients who then use the private key to decrypt the model. Although this methodology ensures the safe passage of model parameters through encryption, \textcolor{black}{introduction of advanced deciphering techniques have made decryption tasks simpler}. Instead of using encryption, \cite{al2020federated} proposes the use of mimic learning in combination with FL for better security and protection against reverse engineering attacks. In Mimic Learning, a teacher ML model is trained on a labeled private dataset. Then a student ML model is initialized and trained on an unlabelled public dataset, which is labeled by using the teacher model. Using this architecture, the probability of exposure of student model parameters reduces drastically. The work was able to achieve 98.11\% accuracy using the NSL-KDD dataset with minimized security risks.

\subsection{FL for IDS in low power IoT devices}

Low power IoT devices in client networks, especially edge devices are major targets of data-hungry attackers and snoopers. \textcolor{black}{ Be it high-end devices that are used in smart IoT like laptops and cameras, or low consumption gadgets like wearable watches and glasses, the data collected from them is confidential and sensitive making it a huge target.} Due to the low power allocation by design, the processing capability of the device is limited. \textcolor{black}{Limited power availability automatically results in applicability of low security measures}. FL provides a helping hand in this aspect. It utilizes the distributed architecture of the network to exploit the overall computation capability and provide a stronger and robust IDS against intruders.

Interception and meticulous processing of the packets can potentially reveal anomalous behavior of devices in the network. An automated detection mechanism in this regard can help drop the suspicious packets and safeguard the device from suspicious network packets. \cite{nguyen2019diot} introduces a novel system, DÏoT, which is a federated self-learning anomaly detection system for IoT devices. This system uses device-type-specific communication profiles without relying on human interaction and a pre-labeled dataset to train the model. In this approach, the data packets are represented as symbols in a predefined language. This representation enables the use of language analysis techniques to analyze the data packets and effectively differentiate anomalous packets from the stream of normal network packets. In this federated architecture, the anomaly detection system resides at each communication node in the network and the model aggregation takes place in a predefined server. In each anomaly detection system, the network packets entering the device are first captured and the characteristic features of the packets are extracted. These characteristic features are then mapped to the symbols of the language. These symbols are then passed through a probability computation model incorporating a recurring unit (here, GRU). The output parameters are evaluated and the packets are finally classified as Anomalous or Normal. This system has been tested against the infamous Mirai attack. Due to the self-learning nature of the system, the effectiveness of the anomaly detection system increases with time. The system adapts to specific device behavior in its due learning process; the rate of learning is high till about 2.5 to 3 hours and gradually plateaus to provide a false positive rate of 0 at the 25$^{th}$ hour. This system demonstrates the capability of the Federated Architecture to learn from crowd-sourced data in the network and provide privacy to individual clients. DÏOT provides 95.6\% accuracy in the detection of anomalies in 257 ms on average; thus making it suitable for low-power IoT devices and sensor networks relying on the speed of the IDS. In addition to that, it provides exceptionally low false alarm rates in real-time scenarios.

% low security measures
% low computation capability
% Edge Computing
%Cite - \cite{chen2020intrusion} cited under data at rest

%sagnik
\subsection{FL based IDS for various Computer Networks}

Based on the scale of a network of connected devices, networks have been classified into PAN, LAN, MAN, WAN, and satellite networks \textcolor{black}{\cite{5963752}}. These types of networks depend on the number of users, density, distance, and traffic. Therefore a FL framework for IDS needs to be tailored to the scenario, the type of data expected, the amount of generated data, and client device interaction. These factors have a high correlation with the type of attack expected and its scale.

In the recent past, satellite communication has opened up the doors to extraterrestrial and worldwide network. This has evolved as a medium for information and and also for attackers to access and attack any connected device regardless of the distance between them. This makes IDS an essential service for STINs \textcolor{black}{(Satellite-Terrestrial Integrated Networks)}. As discussed in \cite{li2020distributed}, there is a huge gap in the computation resources, bandwidth, and dedicated energy between the satellite networks and the terrestrial networks. It suggests the fact that efficiency in resource usage and useful work done needs to be maximized. \cite{li2020distributed} introduces a STIN security dataset consisting of attacks on client devices, especially in DoS attacks. These attacks are a mixture of those performed in the satellite network and those performed in the terrestrial network. \cite{li2019end} also proposes an FL adapted STIN algorithm and a satellite network topology optimization algorithm to adapt to the IDS. This prototype has been deployed and it outperforms traditional deep learning and intrusion detection system. A low packet loss rate and lower CPU utilization rate reduce the overall energy usage making the system efficient and practical.

In a local area network setting, the number of users can range from 3 devices to a network of about 60 devices. This type of network is typically found in organizations like offices, universities, etc. \cite{sun2020intrusion} and \cite{sun2020adaptive} propose a segmented FL architecture, where instead of implementing a single server model for the federated architecture, multiple global models are assigned. This tackles the problem of non-IID data by grouping the evaluations of client models. Grouping of the models reduces the variance of the weight updates within a particular group thus promoting better model aggregation results and prediction accuracy. A dynamic metric to determine the similarity between the client models has been defined based on which the models are grouped. Careful optimization of the hyper-parameters of the models and the metrics have yielded better performances as compared to generic Federated Learning architectures.

\begin{table*}[h!]
\centering
\caption{Summary of FL literature review for IDS}
\label{tab:my-table}
\resizebox{\textwidth}{!}{%
\begin{tabular}{|l|l|p{3cm}|p{5cm}|p{5cm}|}
\hline
\multicolumn{1}{|c|}{Ref.} &
  \multicolumn{1}{c|}{Topic} &
  \multicolumn{1}{c|}{Dataset} &
  \multicolumn{1}{c|}{Contribution} &
  \multicolumn{1}{c|}{Shortcomings} \\ \hline
\cite{rey2021federated} &
  \multirow{3}{*}{FL   architecture for deployment of IDS} &
  NSL-KDD &
  * FL framework for both MLP and   autoencoder &
  * Computation power of the   server \\ \cline{1-1} \cline{3-5} 
\cite{rahman2020internet} &
   &
  - &
  * Performance for   various architectures of FL (distributed, centralized, decentralized) &
  *   Division of work at the server \\ \cline{1-1} \cline{3-5} 
\cite{8284655} &
   &
  - &
  * Distributed GANs &
  * Latency due to   overhead operations \\[0.3cm] \hline
\cite{li2020enhancing} &
  \multirow{3}{*}{FL for Heterogeneous Anomaly Detection} &
  DARPA &
  * Ensemble type   FL &
  * Traffic   masking \\ \cline{1-1} \cline{3-5} 
\cite{briggs2020federated} &
   &
  MNIST, FEMNIST &
  *   Semi-supervised training &
  * Network   Steganography \\ \cline{1-1} \cline{3-5} 
\cite{zhao2019multi} &
   &
  - &
  * Multi task deep neural neworks for FL &
  * High False Positive rates \\ 
  \hline
  \cite{li2020fleam} & \multirow{1}{*}{FL for DDoS attack Detection} &
  - & Combines FL and fog computing to reduce mitigation time and improve detection
accuracy & Only the DDoS is considered\\
  \hline
\cite{vishal2018soaice} &
  \multirow{4}{*}{Fl for privacy preserving   data handling} &
  - &
  * Deep Reinforcement   learning for ruling out privacy vulnerable clients &
  * Model reverse   Engineering \\ \cline{1-1} \cline{3-5} 
\cite{wang2021towards} &
   &
  - &
  *   Attention GRUs for FL &
  * Back Door Attacks \\ \cline{1-1} \cline{3-5} 
\cite{cetin2019federated} &
   &
  AWID &
  *   Encryption using cryptosystems &
  * Data loss due to   lossy compression \\ \cline{1-1} \cline{3-5} 
\cite{li2020deepfed} &
   &
  CPS Data &
   &
  * Vulnerable   cryptosystems \\ \hline
\cite{nguyen2019diot} &
  FL for IDS in low   power IoT devices &
  IoT Activity,   Deployment, Attack Data &
  * Self- learning   Anomaly Detection System for IoT devices. &
  * Support of low   Security Measures.
  
  * Low computation power
  
  * High Communication Latency \\ \hline
\cite{li2020distributed} &
  \multirow{3}{*}{Fl based IDS for various   Computer Network} &
  TER20, SAT20 &
  * Network Topology   Optimization for Satelite Networks using FL &
  * High Rate of False   alarms \\ \cline{1-1} \cline{3-5} 
\cite{sun2020intrusion} &
   &
  LAN Data &
  *   Segmented Federated Learning &
  * High Energy   Consumption \\ \cline{1-1} \cline{3-5} 
\cite{sun2020adaptive} &
   &
  Distributed Network   Data &
   &
   \\ \hline
\end{tabular}%
\label{tab:review}
}
\end{table*}

\section{Challenges}
\label{chall}

In this section, the limitations observed for FL-based IDS are explored. Although FL has shown superiority over conventional DL models for the deployment of IDS, it has its associated drawbacks and vulnerabilities~\cite{bouacida2021vulnerabilities}. Starting from the broadcast of global models to the transmission of trained models, FL for IDS is susceptible to high latency, false alarms, poisoning attacks, etc. This section thus discusses the key challenges and issues of using FL for IDS.

\subsection{Communication Overhead}

The primary limitation of any FL application is the cost of communication per round of training \textcolor{black}{\cite{asad2021evaluating}}. Federated training demands the communication of models parameters (i.e., w$^0$ $\gets$ weights, $\eta \gets$ learning rate, B $\gets$ batch size, etc) first from the central server to $n$ clients and the trained models from each i$_{th}$ client, back to the server. The server traffic, packet transmission loss, the time taken to communicate the parameters may vary greatly depending on the network bandwidth. Also, the use of different devices in a network entails that the computation capability of each is different. Considering all such aspects, it is coherent that in practical situations the overall throughput of any federated network would be low. In scenarios of using the same in intrusion detection applications, the necessity for efficiency and speed increases even more. As an example, a network with millions of devices that collect, generate, and transmit data continuously could be considered. In such case, even if a client ratio of 1\% is taken, the server needs to handle a traffic load of at least $10,000$ clients every broadcast. In most cases, not only does it have to wait for all clients to transmit their parameters but also aggregates the incoming clients simultaneously. Either way, the resultant server system eventually becomes a bottleneck for the federated architecture. 

%--done

\subsection{Federated Poisoning Attacks}

The superiority of federated architectures comes from the distributed nature of data present in client edge devices. Although this property protects the privacy of the data in transit and avoids their collection in a central location, the data in question is still at risk. In the device of a client (i.e. the source of FL data), the labels of the data can easily be modified. These attacks are called poisoning attacks. If a client pretends to be a benign participant, they are capable of modifying the local and global models to produce poisoned predictions. In the worst case, the global model either collapses during training or shows a false performance against the integrity of the data it was trained on. \cite{zhang2019poisoning} conduct poisoning attacks on FL architecture through the use of Generative Adversarial Networks (GAN). In this proposition, a GAN model is deployed to be trained upon and eventually the private data is cloned. Once trained, labels are poisoned for the data generated by the model. As more and more poisoned data is generated from the client, the amount of effect on the global modes increases. A similar work proposed by~\cite{nguyen2020poisoning} introduces a novel poisoning attack for FL-based IoT intrusion detection systems. It inserts small amounts of malicious traffic in the training data to slowly create a backdoor resulting in poisoning attacks. The work proves the significance of the damage that can be caused by these attacks.

%-->done

\subsection{High False Alarms through the use of non-IID data}

The characteristics of flowing traffic cannot simply be determined by studying some samples. Although its properties are discrete, numerous attributes are needed to analyze network traffic, many of which may play a significant role in some cases and maybe negligible in others. The complexity of intrusion detection data itself makes the task of classification tougher. On one hand, due to the abundance of data, utilization of FL is advantageous while on the other hand, the variance in the large data itself may cause improper training of the local and global models, causing a high number of false alarms. This variance is not only due to the heterogeneity of data sources but also the non-IID nature of the data. In each edge device, the size, type, ambiguity, and complexity of data vary. This results in asymmetrically trained local models. On aggregation, these models with high variance would result in the under-fitting of the global model. In addition to this, if a local client holds data that contains redundant, imbalanced, or poisoned features or values, then the respective model may even fail irrespective of the quantity of training \textcolor{black}{\cite{ma2021federated}}. The failure of one local model may not affect the system, but considering the practicality of real-time situations, the probability of multiple client failures is significant.

%-->done

\subsection{Resource Management in Low Power IoT Devices}

FL has the potential to utilize low-power IoT edge devices (i.e., low-power micro-controllers, Remote Telemetry Units, sensor nodes, etc.) for training \textcolor{black}{\cite{dinh2019federated}}. In reality, the majority of these IoT devices do not have sustainable computation power for parallel and continuous training of DL models. These low-power IoT devices hardly support programmed algorithms or even sufficient security measures. Even though this causes them to require IDS, even more, training the DL model locally may not be possible for most devices of this category. If utilized for training, they can drain out very quickly, or train slowly creating a huge latency in the server. In addition, a complete FL edge device procedure would entail data elicitation, data storage, model training, and communication apart from its regular actions. In worst-case scenarios, multiple devices could fail resulting in a deadlock on the server. Robust energy-efficient algorithms must be devised for the implementation of FL in low-power devices. 

The additional disadvantage of using low power or passive devices is its ability to only support low-power communication protocols like 6LoWPAN (IPv6 over Low-Power Wireless Personal Area Network), LPWAN (Low-Power Wide Area Network) and may not support high-end communication protocols like Wifi, bluetooth, 5G, etc. Irrespective of the training time, the communication latency caused would also be high. Training on such devices, especially for Wireless Sensor Networks (WSN), would also encompass several natural and unpredictable situations. The situations arise from being located in harsh industrial and natural conditions \textcolor{black}{\cite{sunny2021low}}. The data collected by them is essential but is still not feasible to completely extract even through FL.

%-->done

% sagnik
\subsection{Vulnerabilities in Intrusion Detection Setups}
Intrusion Detection Systems may typically be installed as the first line of defense for a single workstation or a computer network. An FL architecture primarily employs DL algorithms to classify or cluster the abnormal properties of the network packets to raise an alert. This leaves the IDS susceptible to a wide array of vulnerabilities. These vulnerabilities can be deemed fatal and threatening to the Confidentiality and Integrity of the system under protection. As discussed in~\cite{warzynski2018intrusion}, the models that the DL models have been trained upon might not be familiar with that particular type of anomaly. Thus the detection of such an anomaly is random and can cause damage without any alarm to the respective authorities. This rate of false negatives to previously unseen data poses a threat to a centrally trained model. Moreover, the systems are left vulnerable to a Trojan horse fashion of attack due to the lack of such IDS' ability to handle a new type of anomaly. In case of even slightest deviation from the labeled anomaly, it could be treated as benign \textcolor{black}{\cite{thakkar2021review}}. At the same time, labeling all the traffic requires huge effort of human annotators sometimes with a specific domain of expertise.

%-->done

% sagnik
\subsection{Precise Deployment of IDS}
Deep Learning models employ an enormous amount of parameters and hyper-parameters. Some of these parameters and hyper-parameters are user-defined and others are trainable. These user-defined parameters include learning rate, $\gamma$, number of hidden layers, dimensions of hidden layers, number of epochs per client model, initial weights, etc. Some of these parameters are model parameters while the others are training parameters. In a federated learning scenario, each client has its own model and the fine-tuning of each of these parameters is not feasible. Due to the heterogeneity and the unpredictable nature of the client data, common parameters for each model would not yield an optimal result. Furthermore, in the case of Intrusion Detection Systems, the data is highly unpredictable based on the type of attacks, source, purpose, and various other factors\textcolor{black}{\cite{begli2021multiagent}}.

%[ons]

% solm
%parameter tuning

\section{Future Directions}
\label{FD}

After a critical review of FL-IDS systems and their challenges in the previous sections, the future scope of development in the field is discussed. Covering those issues, directions and possible solutions are provided for each. Considering the current literature and technological advancement on FL systems, many state of the art solutions seem feasible and applicable to the domain of IDS. Plausible solutions include the incorporation of modern communication protocols, encryption standards, blockchain, lightweight DL models, etc. A summary of the challenges faced, and their solutions proposed are presented in table \ref{tab:challanges}.

\subsection{Communication Efficient Federated IDS}

%Soln:
% Binary Neural Network
%Iot defender 5g  
% Asynchronus FL

Important factors contributing to the communication overhead of a federated architecture are the transmission cost of bulky model parameters and the lag in packet transmissions. The more complex an application the bulkier the resultant global model would be. \textcolor{black}{Transferring them to and from all clients for each round definitely consumes a lot of energy}. To overcome this limitation various compression and encryption standards can be utilized. More importantly, lightweight DL technology can be used for reducing the size of parameters that need to be transferred as well as improve their prediction efficiency. One such algorithm named Binarized Neural Networks (BNN) is proposed by Qin \etal~\cite{qin2020line}. In this work, a BNN is trained using federated architecture for the deployment of IDS. In a BNN, the floating-point weight of a neural network is converted to binary formats containing 1's and 0's. It converts all heavy-duty computations into bit-wise operations. This methodology not only enables efficient communication but also ensures line-speed packet processing and traffic classification. Through the use of the P4 programming language, it demonstrates high classification accuracy as well as a low rate of false alarms \textcolor{black}{\cite{qin2020line}}. 

The other major cause of communication overhead being the absence of efficient communication protocols. The bandwidth range for each communication protocol varies greatly. For example, Bluetooth transmissions under 802.11 radio protocols can reach about 24Mbps where Wifi network bandwidths under 802.11ac can reach 1.7 Gbps. Similarly, a device employing low power protocols like LPWAN can only transmit about 400Kbps. To make FL-based IDS more feasible, the implementation of modern communication technologies is mandatory. In proposition, \cite{gupta2017bandwidth} implementned a IDS using 5G Wireless Communication Network (WCN) for bandwidth spoofing attacks. They analyzed such attacks using game theory while optimizing power and resources. Another example of 5G based IDS is contributed in~\cite{fan2020iotdefender}. This literature introduces a new IDS framework called IoTDefender using 5G IoT and Transfer Learning. It is a one of a kind algorithm that integrates the reliability and connectivity of 5G networks with privacy preserving architecture of federated learning. It also utilizes transfer learning for providing customized learning for better detection models. They perform experiments using public CICIDS dataset as well as private datasets. The obtained accuracy is at 91.93\% in addition to better reliability and privacy.

Apart from communication cost, latency in each round is also caused by traffic accumulation on the server-side. On the basis of security as well as sustainability, this traffic can be harmful to the federated system. The server at times may become overloaded with continuous traffic flow leaving it vulnerable to unknown intrusion attacks. To reduce such traffic accumulation and ensure continuous flow, Asynchronous Federated Learning (AFL) has been proposed in recent literature. In~\cite{9378161}, ASO-Fed, an asynchronous online federated learning approach is proposed. In this algorithm, the central server gets updated in an asynchronous manner to remove system lag. The output is measured against an image dataset and three real-world datasets containing non-IID data. Through this, faster model convergence and a much higher system throughput is attained. Lu \etal~\cite{9022982} proposed a similar approach that involves AFL and self-adaptive threshold gradient compression for continuous and smooth communication in edge devices. They are able to implement this algorithm without the asynchronicity affecting the output performance of the models. \textcolor{black}{Lastly, communication sparsification \cite{ozfatura2021time} has also been proposed to counter accumulating traffic on the server side. In this algorithm, certain clients are selected from the total broadcast. This selection may be done in terms of training efficiency, data quality and overall contribution. Never the less, the final accuracy as well as throughput of the system is increased.}

%-->done

%sagnik
\subsection{Encryption Standards for Federated Learning}
The major concern of FL is privacy. The transfer of the information about the data through weights and their subsequent update in the server model is a feature of FL. This transfer of weights does not contain the user data rather, they incorporate the features of this data, thus protecting the user's private information. These weights however could represent some features of the data that are sensitive and could cause harm if extracted by a third party. Moreover, the weights could contain parameters in the order of $10^5$ to $10^9$ or even higher than that depending on the type of the model, the shape of the inputs, outputs and many other factors. Encryption of the weights comes in handy to tackle these issues. Certain types of lossy and lossless encryption functions could help in the compression of the weights. Although there might be certain information loss, the bandwidth and the speed of data transfer are increased significantly improving the overall throughput and the performance of the deep learning model\cite{zhang2020batchcrypt, li2019end}.

\cite{li2019end} proposed an end-to-end encrypted neural network (ENN) for the transmission of gradient updates in FL. ENN compresses and encrypts the gradient updates in the network simultaneously to provide a higher degree of confidentiality and lower charges for the transmission of the gradient updates for the weights. In this proposed framework, the clients use an encoding sub-network on the gradient updates of the model obtained while training to convert it into a lower-dimensional vector. These low-dimensional vectors are then transmitted to the server. Similar to the clients, a decoding sub-network is deployed on the server to retrieve the gradient updates for the weights to be aggregated to the server model. The encoder and the decoder parameters in the structure of ENN are predetermined and do not cause an overhead load on the communication line. The major distinguishing feature of the ENN architecture is that the encrypted weights need not be explicitly updated for the weights to be aggregated. This makes sure that the gradient updates are not revealed on either side of the communication link. Even if the communication is intercepted the extraction of information from the obtained low dimensional vectors is inherently infeasible. This provides strong confidentiality to the client data and reduces the communication overhead.

% intra federated encryption

%-->done

\subsection{Edge Computing in FL based IDS}

%solns of resource management
% light weight models, tflite

Edge devices are not limited to our smartphones or laptops but extend to low-power IoT devices, sensors, etc. The processing power of each is different through which their computation support for federated architectures is reflected \cite{prabadevi2021toward}. As an example, a high-end smartphone accomplishes training of 5 rounds in 5 units of time. At this same time, a low-end IoT device may only manage 1 or 2 rounds of training. This computational difference causes intra-structural latency. To avoid this, efficient resource management is desired in these low power edge devices. Efficient resource management and optimization algorithms have been proposed for faster and reliable edge computation using DL models.

\cite{ren2019federated} proposed a Deep Reinforcement Learning (DRL) based computation offloading optimization for IoT edge devices. Multiple DRL agents are trained and deployed using FL to create a dynamic IoT system with economical allocation of communication and computing resources. Similarly, in~\cite{9154285} a resource management scheme is developed that balances training time and test accuracy for stabilized overall performance. It emphasizes on data importance, communication, and computation cost to model an ideal resource management scheme. They are able to produce a 4x - 10x reduction in training time while maintaining high accuracy. 

A new practice of implementing lightweight DL models has also uplifted the utilization of edge computing. In~\cite{8984222}, Doriguzzi \etal proposed a Lightweight, Usable Convolution Neural Networks (CNN) in DDoS Detection (LUCID). LUCID is a lightweight CNN developed in the work to classify benign traffic from DDoS attacks. They exploit the pattern recognition ability of CNN instead of using time-consuming feature engineering on ML algorithms for lightweight implementation. Another approach to attack detection in the IIoT using a lightweight model is provided in~\cite{latif2020novel}. The work proposes a lightweight Random Neural Network (RaNN) based DL model for the classification of cybersecurity attacks. RaNNs are modified versions of ANNs that depict biological neural networks and their transmissions in humans more closely. Their predictive capabilities and distributive nature make them more suitable for resource-constrained deployments as in the case of IIoT devices. Through this approach, they were able to achieve an accuracy of about 99.20\% with a quick prediction time of 34.51 milliseconds.

\subsection{Implementation of FL through secure channels}

%soln 
% Cite - \cite{alkadi2020deep}
%digital twin
%blockchain

Intrusion detection systems are placed to prevent cyber attacks on the targeted systems. Deploying such IDS systems using FL increases the possibility of intrusions despite the better performance of the overall system. This is due to the creation of various vulnerable points on becoming a huge network. This is to say that no defense mechanism is full proof and as FL is integrated, the system is not only at risk from a single network channel or a virtual machine but an entire network consisting of the server and numerous clients. To overcome these, collaborative learning has adopted the use of secure communication and transaction channels like blockchain. Blockchain has emerged as a secure way of storing transactions or recording information that makes it almost impossible to hack or intrude. In~\cite{alkadi2020deep}, Alkadi \etal propose a Deep Blockchain Framework (DBF) using collaborative intrusion detection. They use a Bidirectional Long Short Term Memory (BiLSTM) DL model for training the IDS. To make data sharing simpler in the cloud, blockchain is used to store raw data alerts created by participating IDS nodes in the form of secure transactions. The algorithm is not only able to secure data from poisoning and inference attacks but also achieve good accuracy in UNSW-NB15~\cite{7348942} and BoT-IoT~\cite{r7v2-x988-19} datasts. To protect federated architectures from poisoning attacks, \cite{9202584} proposed a blockchain-based network that acts as the central server in this scenario. Each model update is done through blockchain transactions. By utilizing these secure transaction records, selective model aggregation is done based on improvement. The overall system is secured from poisoning attacks not only through blockchain but also through verifiable updates.

An emerging technology that can be used for securing FL-based IDS systems is the digital twin technology~\cite{fuller2020digital}. A digital twin represents a virtual replica of real-world physical objects and systems. A digital twin simulation takes in real-world data of those objects as inputs and predicts the outputs. They are able to predict any vulnerabilities and exploitation that can be used against the system before real deployment. \cite{xu2021digital} proposed a new algorithm for the implementation of digital twin technology for anomaly detection in cyber-physical systems. They used a GCN-LSTM based GAN network for mimicking the cyber-physical systems. The generator of the GAN helps to generate real-world unlabelled data while a Timed Automation Machine (TAM) labels these data. The discriminator is the model that classifies the data as benign or anomalous. Anomaly detection with digital twin (ATTAIN) shows an average increase of 8.39\% increase in prediction accuracy with respect to ADS.  Similarly, other algorithms using digital twin have been developed to support vulnerabilities produced by the massive structure of FL.

% sagnik
\subsection{Optimization of FL and IDS parameters}
Federated Learning majorly employ deep learning models that utilize a wide array of trainable and user defined parameters. Moreover Intrusion Detection Systems are highly sensitive to these parameters. Optimizing these parameters has a direct relationship with better outputs and optimized training. Efforts to optimize these parameters have been made with the major focus on meta-heuristic optimization algorithms like evolutionary algorithms, nature-inspired algorithms, etc.

A multi-objective evolutionary algorithm is employed by~\cite{zhu2019multi} to optimize the connectivity and hyper-parameters of the artificial neural network thus simultaneously minimizing the communication costs and maximize the overall accuracy of the models. \cite{zhu2019multi} used an elitist non-dominated sorting genetic algorithm as the multi-objective evolutionary algorithm. In this algorithm, a parent population is first generated. From this parent population, an offspring population is generated using mutation and crossover functions. A combination of the parent and the offspring population is sorted in a non-dominated fashion into a number of non-dominated fronts based on their dominance relationships. The crowding distance is then calculated for each individual within its front. The parent population for the next generation is then determined by selecting the better solutions front by front. The sorting criteria are directed towards increasing the learning accuracy and decreasing the communication overhead. %\cite{zhu2019multi} 
They also proposed a Modified SET algorithm to construct a sparse neural network. This helps in reducing the total space taken by the parameters thus increasing the bandwidth and decreasing the communication costs.
%. can we put the G. and H. subsection of challenges section in the future direction?? (ons) --> they do seem like topics for future directions. let's discuss in the meeting in the evening and place them accordingly. --> ok

%sagnik
\subsection{Efficient handling of non-IID data}
Federated Learning provides a personalized prediction for each client based on the experience from the data generated and usage by the clients. The data generated by the users in the real-time scenario is not homogeneous and is subject to high amounts of noise. Specifically, in the case of IDS, the properties and characteristics of the intercepted packets vary greatly based on usage, time, purpose, and many other factors. Thus it cannot be projected that the data is uniform, identical or independent of one another. These properties of the data tend to cause irregularity and issues in the training and aggregation phases of the client models. Thus handling of the non-IID data becomes an essential part of FL frameworks to ensure proper training and fault tolerance.

One of the techniques to handle non-IID data as proposed in~\cite{briggs2020federated} is to group the client models into smaller clusters based on the similarity of weight updates. This method of performing hierarchical clustering before training and aggregation attempt to preserve uniformity in the cluster. A dedicated cluster head aggregates the weight updates in its locality and acts as a server for the federated framework within the cluster. The cluster heads then act as the clients for the next level of hierarchy for the federated architecture. This type of clustering can provide divided uniformity in an otherwise non-uniform data scenario. In~\cite{briggs2020federated}, the weight updates have been used as the parameter for clustering to preserve the confidentiality of the client data. Since the weight updates represent the actual features of the data, it serves as a better parameter over other variables as criteria for clustering. The additional major advantage of hierarchical clustering is that it reduces the workload of the server by distributing this workload with the cluster heads. This increases efficiency in communication and reduces the overall communication cost. \cite{wang2020optimizing} proposed an experience-driven control framework, namely FAVOR. This framework focuses on intelligent device selection for aggregation in federated learning architecture. FAVOR relies on reinforcement learning to create a smart agent that selects a subset of the devices from the received weight updates during a round. This behavior of the agent counteracts the irregularity in the training caused by the non-IID data. The agent is designed and trained to maximize the accuracy of the server model at the end of every round. The filtration of the misleading weight updates trains the server model properly and converges quickly with better performance of the global model.

% Soln
% Federated Transfer Learning
%Soln:
%Cite - \cite{schneble2019attack
%Cite - \cite{hei2020trusted}

%--> [ons] to improve...

\subsection{FL model heterogeneity and interpretability}

Existing FL solutions mainly use the same ML/DL model (i.e., all the clients have the same model and same model architecture), where edge devices are highly heterogeneous in terms of hardware (e.g., CPU, memory). Developing the same ML/DL may not be possible on some devices. Moreover, using different models/architectures could help the FL system to benefit from this heterogeneity and hence can increase its performance. Thus, more attention should be given to this direction and study the performance of FL system with heterogeneity model or architecture \cite{yang2020heterogeneity, pang2020realizing}.

On the other hand, while using FL to make decisions for a specific scenario it is important to gain insights into the decision-making process of the models. In other words, the interpretability of the FL model is of paramount importance. For example, when in case of FL yielding poor performance, it is necessary to understand the source and reason of the underlying problem. Therefore, it is necessary to explain the final results to find the issue or the strength of the FL system.

\begin{table*}
\caption{Summary of Challenges and Future Directions for FL based IDS}
\begin{tabularx}{\textwidth}{| m{0.5cm} | m{4cm} | m{6cm} | m{5.9cm} |}
  \hline
   \textbf{SL No} & \textbf{Challenges} & \textbf{Description} & \textbf{Future Directions}\\ 
   \hline
 1 & Communication Overhead & \begin{itemize}
  \item High Communication Overhead due to transmission of model parameters to and from the clients each round.
  \item Server bottleneck caused over handling of IDS systems based on a huge network of edge devices.
\end{itemize} & \begin{itemize}
  \item Binarized Neural Networks (BNN) are implemented in federated IDS systems for line-speed traffic classification and minimal latency \cite{qin2020line}.
  \item Modern Communication protocols like 5G and 6G are sought for better reliability \cite{fan2020iotdefender}.
  \item Asynchronous Federated Learning (AFL) methodology is proposed that ensures continuous communication and  prevents central server bottleneck \cite{9378161}.
\end{itemize}\\  
 \hline

 2 & Federated Poisoning Attacks & \begin{itemize}
  \item  Manipulation of local client data and generation of poisoned data \cite{zhang2019poisoning}
  \item Plausible creation of backdoors in federated architecture for data poisoning attacks \cite{nguyen2020poisoning}
\end{itemize}  & \begin{itemize}
  \item Use of digital twin technology to ensure that solely intended data updates are used \cite{fuller2020digital}
  \item Using blockchain based network transactions for the federated architecture to maintain secure transaction records. These transaction records assist in selective model aggregation preventing poisoning attacks.
\end{itemize}\\
  \hline
  3 & High False Alarms through the use of non-IID data & \begin{itemize}
  \item High complexity, variance and ambiguity of federated data cause low  convergence rate.
  \item Data redundancy may result in model collapse.
  \item Inefficient training due to above reasons of local and global model results in a high number of false alarms.
\end{itemize} &  \begin{itemize}
  \item Hierarchical clustering techniques to combat the heterogeneity by grouping similar client weight update responses which simulates homogeneity. \cite{briggs2020federated}
  \item Reinforcement learning techniques to fabricate an agent that rejects misleading weight updates due to heterogeneous data properties.
\end{itemize}\\  
 \hline
 4 & Resource Management in Low Power IoT Devices &\begin{itemize}
  \item  In addition to have low support for security measures, low power IoT devices also have low computation and communication capabilities.
  \item Minimal support for high end communication protocols may cause a system lag in the federated training procedure. 
\end{itemize} &  \begin{itemize}
  \item  Efficient Resource Management, allocation and optimization techniques for better throughput and overall edge device performance \cite{ren2019federated}.   
  \item Implementation of lightweight ML/DL models for faster prediction speeds and lower computation costs \cite{latif2020novel}. 
\end{itemize}\\  
 \hline
 5 & Vulnerabilities in Intrusion Detection Setups & \begin{itemize}
  \item  Continuous threat to the integrity of information assets protected by the IDS.
  \item  Difficult to classify new anomalous traffic, especially in an heterogeneous environment.
\end{itemize} &  \begin{itemize}
  \item Utilization and implementation of Blockchain for secure communication between server and clients \cite{9202584}.
  \item Introduction of digital twin technology in FL based IDS systems for better trustworthiness of FL architectures \cite{xu2021digital}. 
\end{itemize}\\  
 \hline
 6 & Precise deployment of IDS & \begin{itemize}
  \item  Huge pool of parameters that need to be fine tuned.
  \item  Slight deviations might not produce optimal results.
  \item  Manual tuning is time consuming and a tedious process.
\end{itemize} &  \begin{itemize}
  \item Utilizing meta-heuristic optimization algorithms to tune the parameters mid-training. \cite{zhu2019multi}.
  \item Reducing the overall number of parameters to provide fault tolerance. 
\end{itemize}\\  
 \hline
\end{tabularx}
\label{tab:challanges}

\end{table*}

\section{Conclusion}
\label{conc}
The enormous success of ML techniques and related applications have been guided by three major factors. Firstly, the availability of data collected through years. Secondly, the evolution of technology enabling enhanced computational power which have led to the development of flexible and integrated devices, that trains data model faster in reduced costs. Thirdly, the use of deep learning models which have enabled self-learning ensuring optimized accuracy. But there exists several challenges associated with data privacy and security. The modern world mobile phones, smart domestic and professional devices, wearables, autonomous vehicles generate humongous amount of data every single day. The privacy protection of such data and related devices is a necessity. The machine learning based Intrusion detection systems (IDS) aims to fulfil that necessity.  But these frameworks still face issues to meet the privacy and security requirements due to its need to store and communicate data to the centralized server. The use of federated learning based IDS solutions train high-quality centralized model ensuring user privacy. The paper provides a comprehensive survey of FL based solutions in IDS emphasizing on the security, privacy and reliability aspects. The paper also discusses the challenges  and vulnerabilities in FL implementations pertinent to high latency, false alarms, poisoning attacks and various others. The negative impact of such issues on various aspects of IDS are also described. The paper finally presents potential scope of future research and also proposes possible solutions to combat the challenges of FL implementations in IDS.

%\balance
\bibliographystyle{plainnat}
\bibliography{references}

% \begin{IEEEbiographynophoto}{Quoc-Viet Pham} [M'18] (vietpq@pusan.ac.kr) is currently working as a research professor at Korean Southeast Center for the 4th Industrial Revolution Leader Education, Pusan National University, Korea. He has been granted the Korea NRF Funding for outstanding young researchers for the term 2019–2023. He received the best PhD thesis award in Engineering from Inje University in 2017. His research interests include network optimization, edge computing, resource allocation, and wireless AI.
% \end{IEEEbiographynophoto}

\end{document}